# Urban Haze and Photovoltaics

I. M. Peters, S. Karthik, L. Haohui, T. Buonassisi, A. Nobre

**Abstract:**

Urban haze is a multifaceted threat. Foremost a major health hazard, it also affects the passage of light through the lower atmosphere. In this paper, we present a study addressing the impact of haze on the performance pf photovoltaic installations in cities. Using long-term, high resolution field data from Delhi and Singapore we derive an empirical relation between reduction in insolation and fine particulate matter (PM2.5) concentration. This approach enables a straightforward way to estimate air pollution related losses to photovoltaic power generation anywhere on the planet. For Delhi, we find that insolation received by silicon PV panels was reduced by 11.5% ± 1.5% or 200 kWh/m$^2$ per year between 2016 and 2017 due to air pollution. We extended this analysis to 16 more cities around the planet and estimated insolation reductions ranging from 2.0% (Singapore) to 9.1% (Beijing). Using spectrum data from Singapore, we projected how other photovoltaic technologies would be affected and found an additional reduction compared to silicon of between 23% relative for GaAs and 42% for a 1.64 eV perovskite material. Considering current installation targets and local prices for electricity, we project that annual losses in revenue from photovoltaic installations could exceed 20 million USD for Delhi alone, indicating that annual economic damage from air pollution to photovoltaic site operators and investors worldwide could be billions of dollars.

**Motivation:**

*In June 2013, three of the authors of this paper, I. M. Peters, L. Haohui and A. Nobre, lived in Singapore and were witnesses of the most severe haze event to have occurred in the city to date. For a couple of days, the pollutant standard index (PSI) jumped from its usual value of about 25 to over 200. The normally clear view from our office windows on the sixth floor became filled with an impenetrable fog that swallowed up neighboring buildings. Face masks were sold out in a matter of hours. People were panicking. The event, in many ways, served as a wake-up call. Given our research focus on photovoltaic installations, we wanted to investigate the impact of haze on solar cell performance. From the reduced visibility it was evident that haze must have an effect, and we set out to quantify it. This paper summarizes our understanding so far. We have since learned about the devastating effects of urban air pollution on human health. This paper adds another aspect – the detrimental effect on photovoltaic power generation due to the reduction of light received. We hope that, in a small way, we can help raise awareness and make progress to improve the quality of life in what more and more people call home in the 21$^{st}$ century – cities.*

**Introduction:**

Air pollution in cities is a problem of growing urgency [1]. According to the World Health Organization (WHO), global urban air pollution levels have increased by 8% between 2008 and 2013 [2]. The highest urban air pollution levels were observed in low- and middle-income cities in the Eastern Mediterranean and Southeast Asia. More than 80% of people living in monitored urban areas were exposed to air quality levels that exceed WHO limits [2].

Air pollution typically includes ozone, small- (PM10) and fine (PM2.5) particular matter [3]. Fine particulate matter describes particles with a diameter of less than 2.5 µm. The primary sources of these particles are incomplete combustion, automobile emission, dust and cooking. In its majority, they are anthropogenic, with major composites being sulfates, nitrates, ammonia, carbon, lead and organic matter. Due to their small size, PM2.5 particles are a major health hazard, because they can enter lungs and bloodstreams of humans. They give rise to chronic damage to the respiratory- and cardiovascular system [4, 5] and contribute to premature mortality rates [6, 7]. WHO estimates that air pollution causes about 6.5 million premature deaths every year [8a].

Apart from devastating effects on health, PM2.5 particles are also the main cause of haze – periods with reduced visibility occurring in urban areas all around the world [8 – 14]. **Figure 1** shows examples of how haze changes visibility in six cities. The change in visibility is related to a reduction in solar intensity and an alteration of the spectrum reaching the ground.

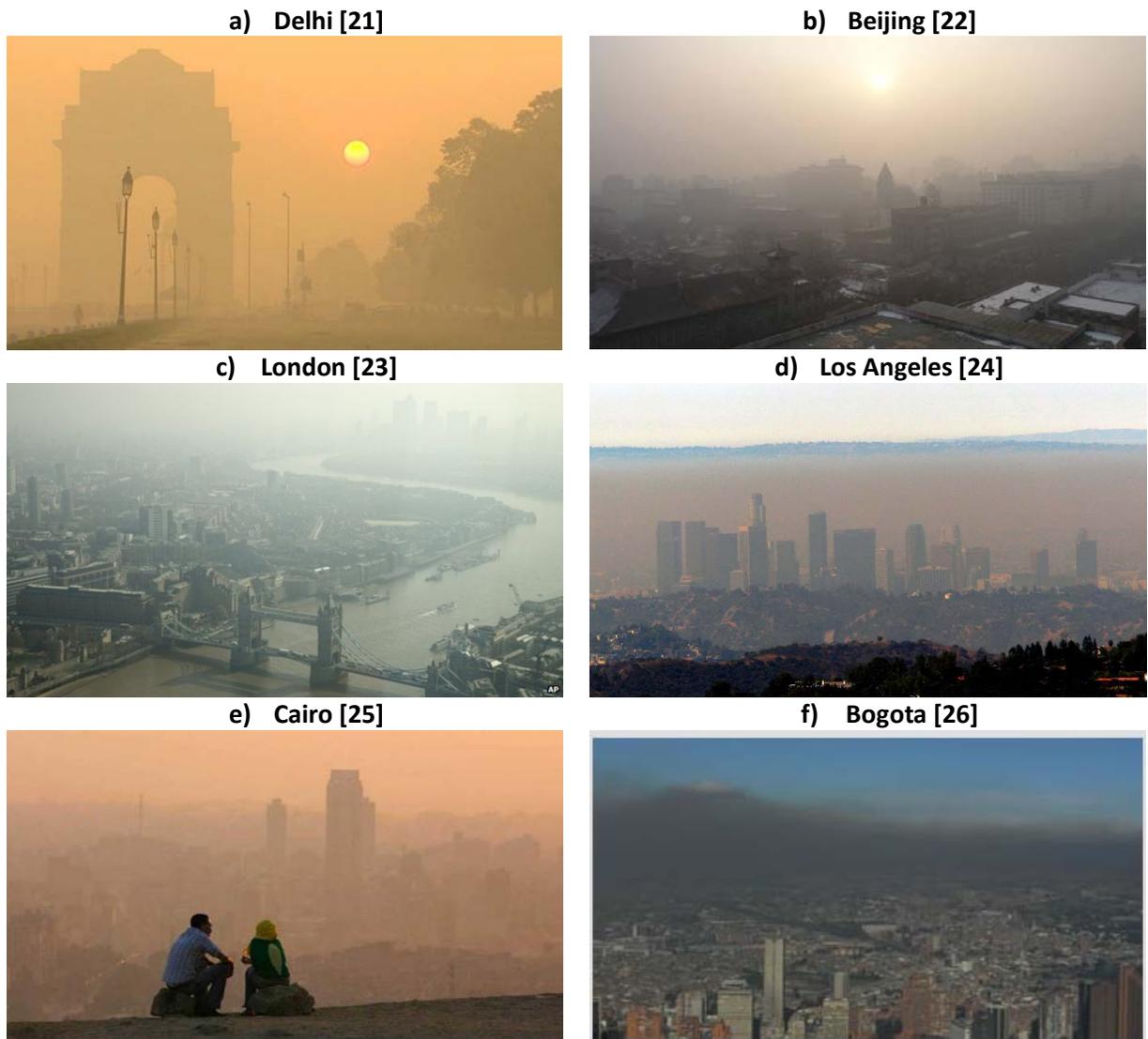

*Figure 1: Haze events around the world.*

In this study we focus on quantifying the impact of haze on insolation levels in cities, and the reduction in power generated by photovoltaic panels due to this effect. Until 2050, it is expected that 2.5 billion people will live in urban areas [15]. City integrated photovoltaics (PV) offer opportunities of mitigating challenges related to the high power demand of these growing urban areas [16, 17].

Surprising to the authors, it was difficult to assess the global capacity of PV installations in cities, based on literature data. However, there are numerous indicators that a significant PV capacity is installed in urban areas, and more is to come. The potential for rooftop solar PV installations in cities was estimated at 5.4 TW – 70% of the electricity demand of urban residential and commercial consumers [17a]. Only a small fraction of this potential has been utilized so far. According to *Environment America*, in 2014, 6.5% of the solar PV capacity in the United States, or 1.3G W, were installed in cities, with Los Angeles in the lead [18]. Several countries have formulated targets for rooftop installations, the most remarkable being India with 40 GW planned by 2022. China already has 22 GW of combined PV capacity installed on rooftops, and a further boost in rooftop installations is expected in 2018 [19]. Australia had 16% of PV capacity installed on rooftops as of 2016. Furthermore, a growing number of cities have committed themselves to achieving 100% renewable electricity or energy. Examples include San Diego and San Francisco (California), Vancouver (Canada), Copenhagen (Denmark), Munich and Frankfurt (Germany). Other cities have formulated targets for renewable shares of total energy – e.g. Austin (Texas) 65% by 2025 and Paris (France) 25% by 2020 – or total electricity – e.g. Amsterdam (Netherlands), 50% by 2040, Canberra (Australia), 90% by 2020 and Tokyo (Japan) 24% by 2024 [20]. While neither rooftop installations nor energy targets translate directly into PV installations in cities, they are indicators for the growing amount of solar panels installed in or near cities. There are also concrete targets for the installation of local renewable electric capacity in cities, including Los Angeles (1.3 GW of solar PV by 2020), New York (350 MW of PV by 2024) and San Francisco (950 MW renewables by 2020) [20].

The goal of this work is to correlate fine particulate matter concentration with a reduction of insolation reaching solar installations in cities. We start by analyzing pollution and insolation data collected in Delhi, India. The availability of high-quality, high-frequency insolation- and pollution data, as well as the high levels of pollution observed in Delhi make this city a suitable test case to observe the effects of air pollution. We apply a filtering method, introduced in [14], to find the functional relation between PM2.5 concentration and relative reduction in insolation. We then use this functional relation to estimate haze related insolation losses in a number of cities around the globe, and to project the impact on different PV technologies. Finally , we attempt to estimate the economic damage resulting from air pollution related reductions in insolation.

**Correlating Fine Particle Matter Concentration and Yield Losses in Delhi:**

I PM2.5 Concentration and Insolation Data

The collected data used in this analysis is summarized in **Figure 2**. On the left hand side (a), measured insolation of a photovoltaic installation in Paschim Vihar, Delhi is shown. Insolation was recorded between May 2016 and November 2017 using two daily cleaned silicon sensors and a pyranometer with a frequency of one measurement every minute. The data shown in the figure was measured with one of the silicon sensors as this data is most representative of a silicon PV installation.

On the right hand side (b), PM2.5 data recorded at the U.S. Embassy and Consulates' air quality monitors in Delhi [27] is depicted. The data is recorded with a frequency of one measurement per hour and made available through the "AirNow:" website of the Department of State [28]. The sensor is located in the Chanakyapuri area, approximately 20 km away from the sensor measuring insolation. In the further analysis we assume that the general trends of PM2.5 concentration are consistent over the area covering both sites. This assumption was confirmed by measurements taken at various sites in Singapore.

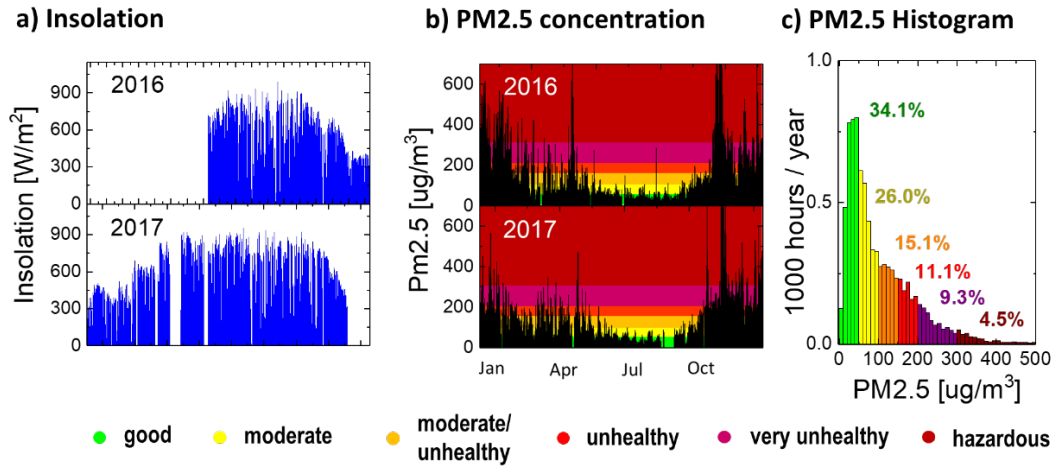

*Figure 2*: a) Delhi insolation measured with a silicon sensor between May 2016 and November 2017. b) Delhi PM2.5 concentration measured by the U.S. Embassy and Consulates' air quality monitors [27, 28] c) histogram of PM2.5 concentration durations.

To classify PM2.5 concentration levels, we follow color coding for Air Quality Index (AQI) as suggested by AirNow [29]. Note, however, that AQI and PM2.5 concentration are related but dissimilar metrics, hence pollution levels are not directly transferrable to AQI levels as typically used. Terminology and color coding is summarized in **Table I**.

*Table I:* PM2.5 Concentration ranges and terminology

| Concentration range ($\mu g/m^3$) | Color Code | Levels of Health Concern |
|---|---|---|
| 0 – 50 | **Green** | good |
| 51 – 100 | **Yellow** | moderate |
| 101 – 150 | **Orange** | moderately unhealthy |
| 151 – 200 | **Red** | unhealthy |
| 201 – 300 | **Purple** | very unhealthy |
| Above 300 | **Maroon** | hazardous |

II Correlating PM2.5 Concentration and loss in Insolation

The methodology for relating PM2.5 concentration to a reduction in insolation was introduced in [14]. For a detailed description, we refer readers to this publication, though we provide a brief summary of the approach in the following. In [14] we suggested sorting insolation data in bins corresponding to different levels of PM2.5 concentration, and then use humidity and clear sky filters to identify data points representative of clear sky conditions.

We generally apply the same method here, yet with some alterations. The main difference between the analysis presented in [14] and here is the length of time period considered. In [14] we analyzed a haze event that took place within 18 days. Here we consider insolation data over 19 months. Over the course of a year insolation varies via the zenith angle and the eccentricity of the Earth's orbit around the sun [30, 31], as well as due to seasonal variations in atmospheric conditions – e.g. water content. These variations need to be considered when analyzing correlations. For the purpose of this study, we considered data for different moths separately. Data of different months was later combined after normalizing it to the 90 percentile of all data points collected at noon for a given month. The effect of this normalization is shown in Figure 3 for the examples of August and December, marking the months with highest and lowest overall insolation.

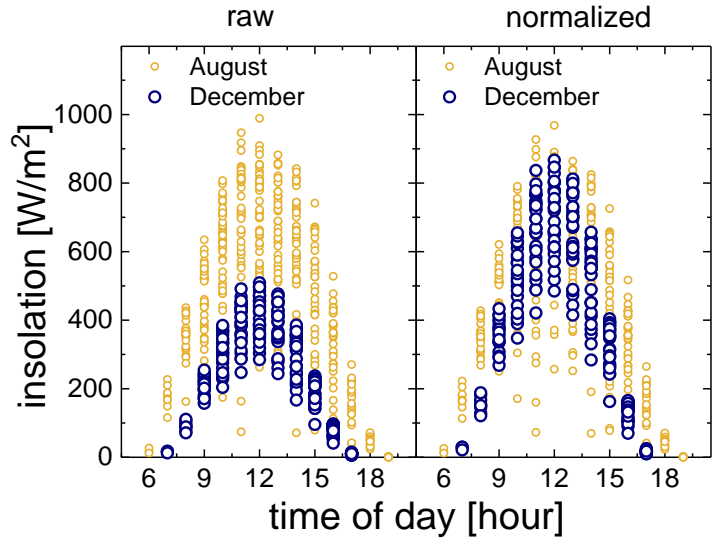

***Figure 3***: *Data measured in the months of August (yellow) and December (purple). On the left side the raw data is shown, on the right we show the data normalized to consider seasonal variations.*

**Figure 4** illustrates how we correlate PM2.5 concentration levels with a reduction in insolation. **Figure 4a** shows in steps how typical daily insolation curves for different pollution conditions are obtained. The figure on the left shows all available data points after adjustment for seasonal variations, with color coding indicating which pollution level they belong to. In the middle, two specific conditions are selected – moderate (50 – 100 µg/m$^3$) and very unhealthy (200 – 300 µg/m$^3$). The lines indicate the 80 percentile of the ensemble for each hour and pollution level. The 80 percentile filter here fulfills the same function as the combined humidity and clear sky filter in [14] – i.e. it identifies conditions that are representative of a clear sky. The picture on the right, finally, shows the curves obtained for all six pollution levels. Note that only ensembles (i.e. data sets for a specific hour and pollution level) were considered that contained more than eight data points. This data was then used to generate **Figure 4 b** by considering the relative decrease in intensity as a function of PM2.5 concentration for each hour. Note that it is possible to correlation PM2.5 concentration and insolation reduction in a variety of different ways. We are discussing some of them in the **supporting material**. The important finding is that all ways we have tried resulted in similar functional correlations i.e. mono-exponential decays with exponents that were not significantly different.

Also shown in **Figure 4 b** are previous results from [14] (cross shapes). These results are consistent with findings presented here. The advantage of the data set acquired in Delhi is the wide range of pollution conditions covered. Data ensembles with statistical relevance could be collected up to 375 µg/m$^3$ concentration, a factor three more than for Singapore. One consequence of this wider range is that the data presented here is slightly better represented by an exponential decay than by a linear function ($R^2$ of 92% compared to 90%). Exponential decay is expected according to Lambert-Beer's law. The fitted exponential decay is shown as a blue line in the figure, and the functional relation is written down in equation 1.

| a) filtering… | b) …and correlating |
|---|---|

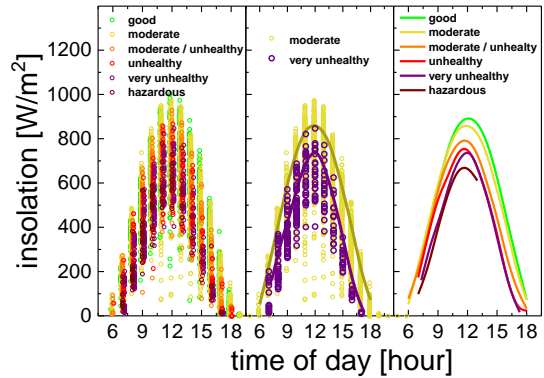 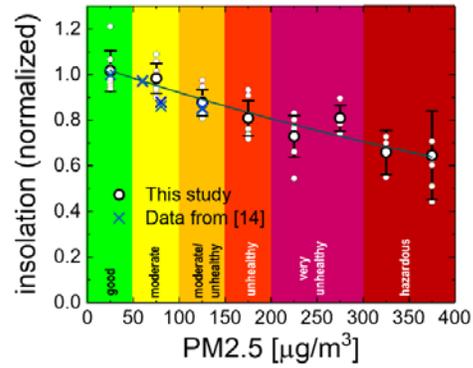

*Figure 4: a) Sketch of how data points and typical curves for a clear day under specific pollution conditions are obtained. The figure on the left shows the data after adjustment for seasonal variations with a color code depending on pollution level. In the middle figure, two conditions are highlighted and curves are shown that represent the 80 percentile for the data ensemble of each hour. The figure on the right shows the curves obtained for all pollution conditions. In b) this data is used to formulate a functional correlation between PM2.5 concentration and reduction of insolation by considering the relative reduction for each hour. Also shown are data obtained previously for Singapore [14] and a mono-exponential decaying fit (see equation 1).*

The function describing the exponential decay in **Figure 4 b** is summarized in the following equation. We will use this equation in the following to estimate how much light is lost due to air pollution. The uncertainty given here is calculated from least mean square fitting of a mono-exponential decay. Further details about this fit are given in the supporting material.

$$\frac{I(PM2.5)}{I_0} = exp\left(\frac{-PM2.5}{750 \pm 90}\right) \quad (1)$$

III Haze related reduction in insolation and impact on a Si PV panel

The relation formulated in **Equation 1** can directly be used to estimate the loss in yield due to a reduction in insolation due to haze. **Figure 5** summarizes the measured insolation for 2016 and 2017 from **Figure 2** in blue, and the estimated insolation at 0 µg/m³ PM2.5 concentration in orange. The estimated curve was obtained by adjusting the measured insolation using the PM2.5 concentration measured at the same time.

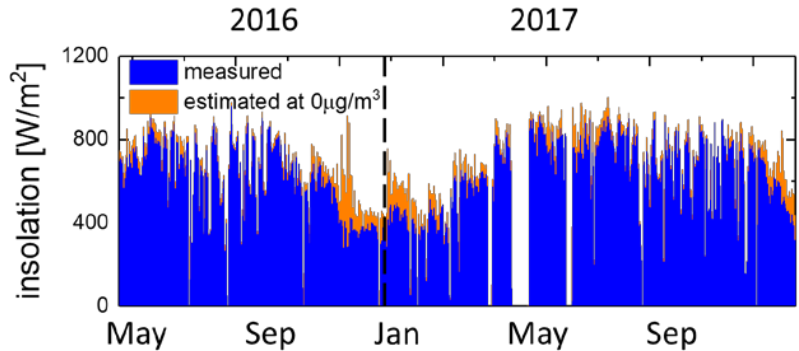

*Figure 5: Measured insolation (blue) and estimated insolation without air pollution, i.e. at a PM2.5 concentration of 0 µg/m³ (orange).*

Integrating the insolation over the period of one year (May 2016 to May 2017), we obtain the annual radiant exposure as well as the projected exposure at 0 µg/m³ PM2.5. These values are summarized in **Table II**. The projected losses for Delhi are 11.5% of the annual solar energy or 200 kWh/m². We have also included the projected yield losses for a 20% efficient silicon PV module.

*Table II:* PM2.5 Impact on insolation in Delhi over one year.

|  | Annual Radiant Exposure [kWh/m²] | Annual yield of a 20% Si-module [kWh/m²] |
|---|---|---|
| Measured data | **1570** | **314** |
| Projection 0 µg/m³ PM2.5 | **1770 ± 30** | **354 ± 6** |
| Ratio | 88.5 ± 1.5 % | |
| Difference | 200 ± 30 | 40 ± 6 |

**Projected losses in other cities:**

Using the functional relation formulated in **Equation 1**, we extended the assessment of loss in annual radiant exposure to other cities. While it can be argued that the composition of haze and its optical behavior vary from location to location [32], we found that the relation between PM2.5 concentration and insolation we observe for Delhi also describes data from Singapore reasonably well (compare **Figure 4b**). We take this as an indication that an estimate based on Delhi data can provide at least some guidance about what losses to expect. The cities covered in this study are shown in **Figure 6**. We selected cities based on availability of air quality data, known haze events and geographic distribution.

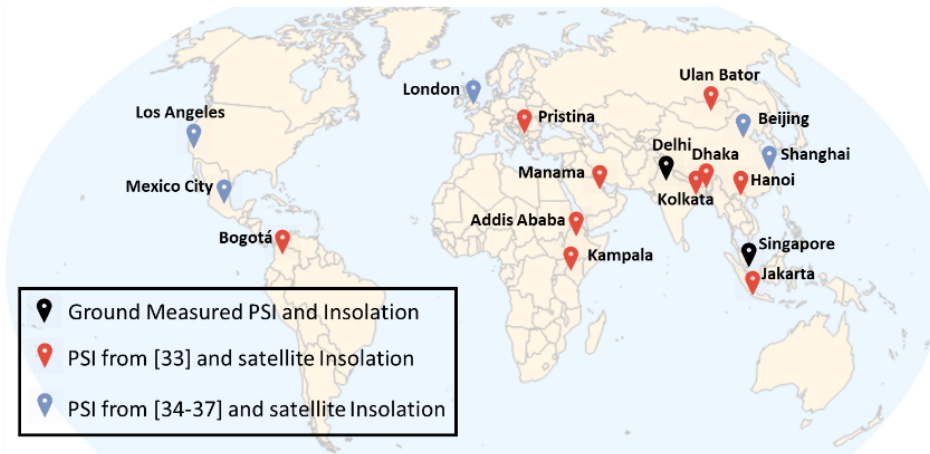

*Figure 6:* Map of cities covered in this study. Black markers indicate locations for which simultaneous, ground-measured PM2.5 and insolation data are available. Red markers indicate locations for which PM2.5 data was obtained from [33], for blue markers PM2.5 data was taken from [34 - 37].

Air quality data is available from a number of sources. For the results presented in this paper, we have used data from the AirNow Department of State (DOS) website [33]. This website provides historical data of PM2.5 measurements for a number of cities measured at U.S. embassies and consulates around the world. Depending on location, historical data dates back as far as 2015. For consistency reasons, we have made use of this database wherever possible. Additional sources we used (marked in blue) come from National or Communal centers and institutions [34 – 37]. The data obtained is summarized in **Figure 7**.

a) PM2.5 time series I.

b) PM2.5 time series II

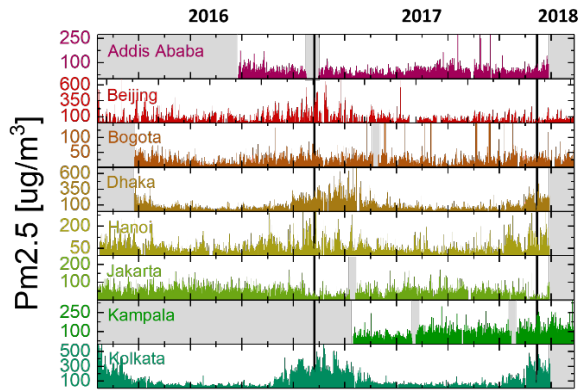
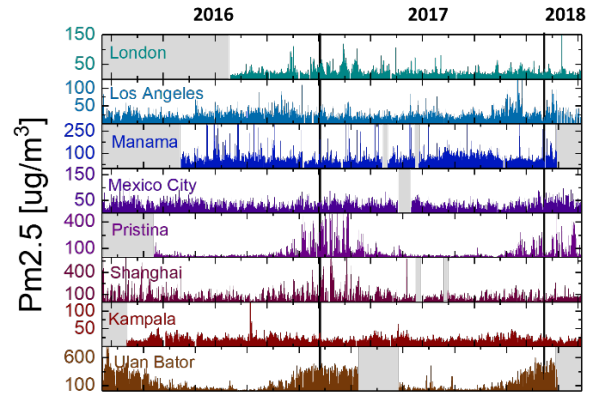

c) PM2.5 histogram I

d) PM2.5 histogram II

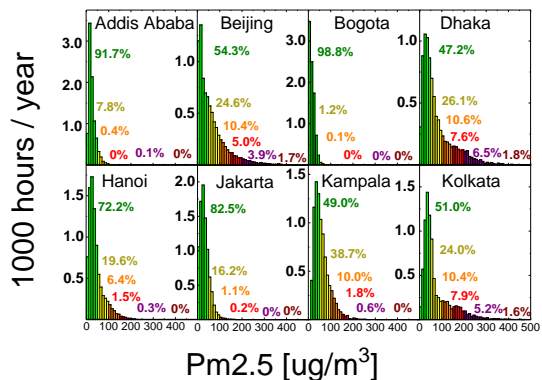
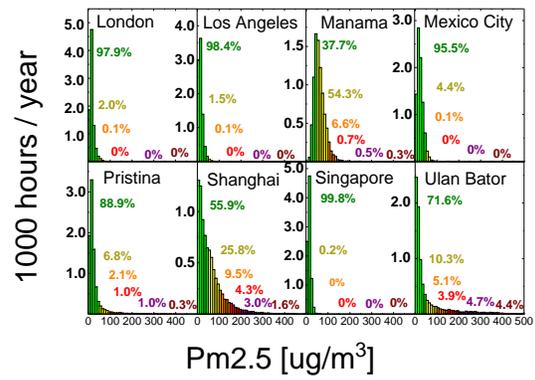

**Figure 7:** Summary of PM2.5 data obtained for cities shown in **Figure 6**. a) and b) show the time series of the data obtained from [33] and [34 -37]. Grey areas mark periods for which no data was available. c) and d) show the corresponding histograms of this data over the course of one year.

Note that recordings for several cities are imperfect. Examples are Bogota in May and June 2017, Moscow in September 2017 and Richards Bay after October 2017. A dataset that does not cover a full year holds the threat of not correctly representing seasonal influences, which results in a hard to predict uncertainty. We have decided to still include the data for these cities, but ask the reader to consider this shortcoming when interpreting the results.

Using the data summarized in **Figure 7**, and equation 1, we can estimate the relative and absolute losses in radiant exposure. This estimate is summarized in **Figure 8** and **Table III**. As we didn't have access to ground measured irradiance data for all locations, we used typical yearly irradiances, obtained from satellite data [38] and a clear sky model [31, 39, 40]. Satellite daily average irradiance data from NASA is available with a 1 x 1 degree resolution. Irradiance for each day of the year over the course of ten years (2005 – 2015) was used, with the median value for each day marking the considered typical value (see supporting material). Daily irradiance distribution was projected with the clear sky model. Corrections with equation 1 were calculated on an hourly basis by reducing the incoming photon flux according to PM2.5 concentration. This approach neglects contributions of clouds as well as ground reflection and scattering. Errors can be expected if such effects occur systematically at a specific time of day, as is the case, for example, in Singapore where morning tend to be sunnier than afternoons [41]. To benchmark this approach, we calculated the relative losses for Delhi (shown in the last column in Table III a). The calculated losses are 12.2 ± 0.9 %, which is consistent with the 11.5 ± 1.5 % obtained from ground based measurements.

To project the impact on photovoltaic panels, we have included two metrics in **Table III**, the global tilted irradiation for fixed systems at optimum angle (GTI) in kWh/m$^2$, and the Photovoltaic power potential ($PV_{out}$) in kWh/kWp. Values for these metrics for the different cities were obtained from the Global Solar Atlas [42]. Corrections and losses are calculated directly by adjusting for the relative loss (first row). Uncertainties were obtained directly through the uncertainty in equation 1 and indirectly by varying the beginning- and end point of the considered one year period.

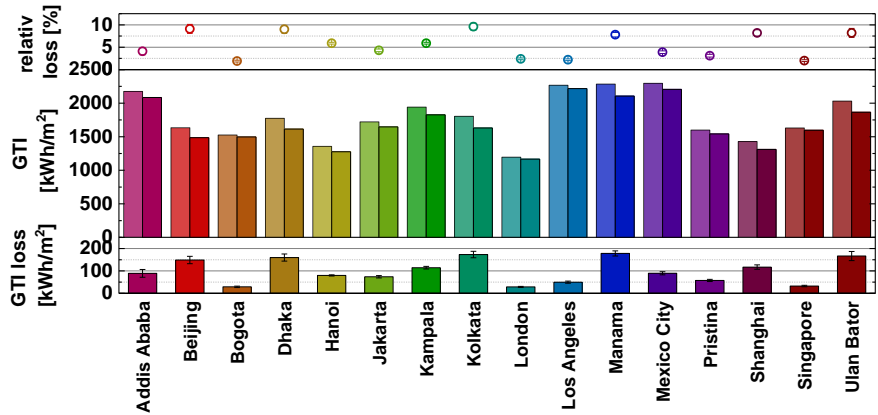

*Figure 8: Projected relative losses in radiant exposure for the considered cities - top, as well as tabulated global tilted irradiation for fixed systems at optimum angle (GTI) values according to Ref [39] (bright colors) and corrected values (dark) - middle. Absolute loss in GTI for each city is shown in the lower part of the figure.*

*Table III a): projected relative losses in radiant exposure for the cities from **Figure 7** a) and c), as well as tabulated and corrected GTI values, GTI losses and losses in PV power potential ($PV_{out}$).*

|  | Addis Ababa | Beijing | Bogota | Dhaka | Hanoi | Jakarta | Kampala | Kolkata |
|---|---|---|---|---|---|---|---|---|
| rel. loss [%] | 4.1±0.8 | 9.1±1.0 | 1.9±0.2 | 9.0±0.9 | 5.9±0.2 | 4.3±0.3 | 5.9±0.3 | 9.6±0.8 |
| GTI [kWh/m$^2$] | 2174 | 1634 | 1526 | 1774 | 1356 | 1721 | 1941 | 1804 |
| GTI corr. [kWh/m$^2$] | 2085 | 1485 | 1497 | 1614 | 1276 | 1647 | 1826 | 1631 |
| GTI loss [kWh/m$^2$] | 89±17 | 149±16 | 29±3 | 160±16 | 80±3 | 74±5 | 114±6 | 173±14 |
| $PV_{out}$ loss [kWh/kWp] | 71±14 | 121±13 | 24±2 | 123±12 | 62±2 | 57±4 | 89± | 132±11 |

*Table III b): projected relative losses in radiant exposure for the cities from Figure 7 b) and d), as well as tabulated and corrected GTI values, GTI losses and losses in PV power potential (PV$_{out}$). Delhi is included as a reference to illustrate differences between ground based data and satellite based data.*

|  | London | Los Angeles | Manama | Mexico City | Pristina | Shang-hai | Singa-pore | Ulan Bator | **Delhi** |
|---|---|---|---|---|---|---|---|---|---|
| *rel. loss [%]* | 2.4±0.2 | 2.2±0.2 | 7.8±0.5 | 3.9±0.3 | 3.6±0.3 | 8.2±079 | 2.0±0.2 | 9.2±1.0 | **12.2±0.9** |
| *GTI [kWh/m$^2$]* | 1195 | 2267 | 2284 | 2295 | 1599 | 1428 | 1630 | 2031 | **1863** |
| *GTI corr. [kWh/m$^2$]* | 1166 | 2217 | 2106 | 2205 | 1541 | 1311 | 1597 | 1864 | **1636** |
| *GTI loss [kWh/m$^2$]* | 29±2 | 50±5 | 178±11 | 90±7 | 58±5 | 117±10 | 33±3 | 167±20 | **227±17** |
| *PV$_{out}$ loss [kWh/kWp]* | 24±2 | 39±4 | 137±9 | 71±5 | 47±4 | 94±8 | 25±3 | 144±18 | **174±13** |

**Projection to other PV technologies:**

Losses in radiant exposure are not distributed uniformly across the solar spectrum. Data acquired in Singapore shows that shorter wavelengths are more affected, which is in line with expectations from basic scattering theory [43]. Hence, solar cells with a larger band gap should be more strongly affected than silicon. We projected the impact on other PV technologies by considering changes in absorption up to the band gap of the specific solar cell technology as a function of PM 2.5 concentration. Four band gaps were considered: silicon (1.12 eV) as baseline, GaAs (1.43 eV), CdTe (1.54 eV) and a material from the perovskite family (1.64 eV) [44, 45] and four locations: Delhi, Beijing, Hanoi and Mexico City. The results of this exercise are shown in **Figure 9** and **Table IV**. Compared to silicon, losses are projected to be increased by 23% for GaAs, 33% for CdTe and 42% for perovskites.

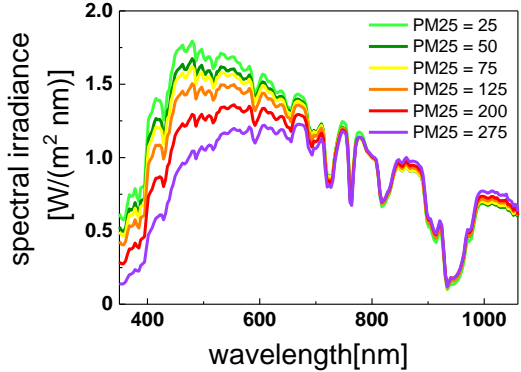
a) Spectrum changes with PM25

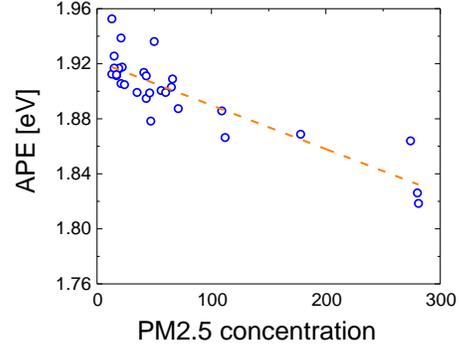
b) PM2.5 concentration vs APE

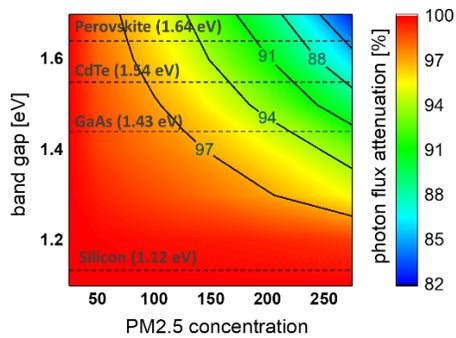
c) projected impact on photon flux

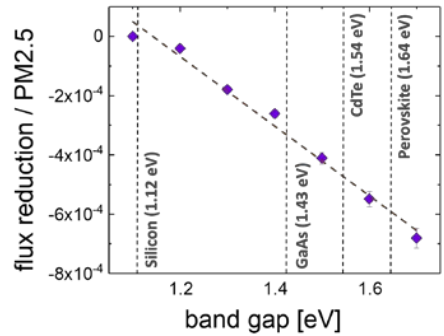
d) linear approximation of flux attenuation

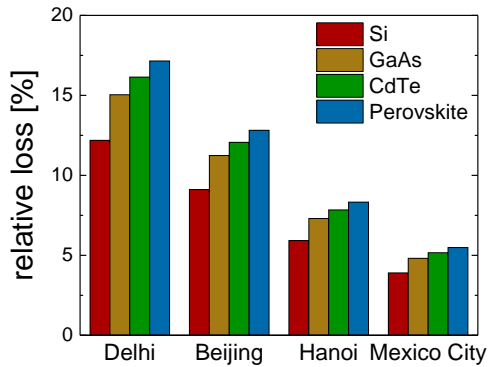
e) projection for cities and technologies

*Figure 9:* Effect of fine particulate matter on different PV technologies. a) Shows on examples how the solar spectrum is affected by different PM2.5 concentrations. b) Summarizes the relation between PM2.5 concentration and average photon energy. The line is a guide to the eye. c) Projects the impact of PM2.5 concentration on the absorbed photon flux for absorbers with different band gaps. d) Illustrates the reduction in photon flux per PM2.5 concentration and band gap. In e) this relation is used to project the PM2.5 induced losses onto different PV technologies.

*Table IV):* projected relative losses in absorbed photon flux for different PV technologies and four exemplary locations (compare Figure 9e)

|  | Silicon (1.12 eV) | GaAs (1.43 eV) | CdTe (1.54 eV) | Perovskite (1.64 eV) |
|---|---|---|---|---|
| **Delhi** | 12.2 | 15.0 | 16.1 | 17.2 |
| **Beijing** | 9.1 | 11.2 | 12.0 | 12.8 |
| **Hanoi** | 5.9 | 7.3 | 7.8 | 8.3 |
| **Mexico City** | 3.9 | 4.8 | 5.2 | 5.5 |

**Economic considerations:**

Using these numbers and Table 3, we can estimate the loss in revenue to PV installers and operators. The average price for residential electricity in Delhi in January 2017 was 12.2 ct/kWh (8.07 ₹/kWh) [46]. The estimated loss in PV power output is 174 kWh/kWp annually, resulting in a loss of 21.23 $US/kWp each year – or 12% relative. Given the targets formulated by the Indian government, rooftop installations in Delhi can be expected to be in the GW range, which would put the annual damage to economy above 20 million dollars. Similarly, the estimated damage in Kolkata is 16 million dollars per GW.

In China, residential power prices are given as 7.9 ct/kWh (0.495 yuan/kWh) in Beijing and 9.1 ct/kWh (0.571 yuan/kWh) in Shanghai [47]. The estimated loss in annual revenue here is 10 million dollars per GW installed for Beijing and 9 million dollars per GW installed in Shanghai.

However, even for cities with better air quality like Los Angeles air pollution causes notable economic damage. Average electricity costs for household electricity in Los Angeles in February 2018 were 18.4 ct/kWh [48] and 11.7 ct/kWh [49] for industry. Using the installation target of 1.3 GW by 2020, annual losses are between 5.9 and 9.3 million dollar.

**Discussion:**

I Comparison to literature

Aerosols affect how sunlight passes through the atmosphere. This has been known since Tyndall observed that blue light is scattered more strongly than red light when passing through a clear fluid holding small particles [50]. And already Tyndall was concerned with the question how small particles that affect the passage of light would influence health. "Nor is the disgust abolished by the reflection that, although we do not see the nastiness, we are churning it in our lungs every hour and minute of our lives." He writes in an article in Nature in 1870 [51]. As discussed earlier, the concerns were well founded and numerous studies have confirmed the detrimental effects of haze on health – even if they are different than what Tyndall imagined.

Attempts to quantify the effect of aerosol particles on solar irradiance have also been made. A study by Husar et al. [52] and Guyemard et al. [53] looked at the effect on solar irradiation caused by two intense dust storms in the Gobi dessert. When reaching the United States, the dust clouds resulted in a 30% reduction in direct normal solar radiation, and a 3.8% to 5.1% reduction in global horizontal irradiance (GHI). The increase in air pollution in China has triggered a number of studies that investigate the correlation between aerosols and solar irradiance. Li et al. [54] correlated satellite measured air pollution data and solar photovoltaic resource in China. They found that the combined aerosol optical depth of all particles reduced solar irradiance by 20% - 35%. Note that this study considers all aerosols in the atmospheric column, whereas results presented here only considers fine particulate matter.

We note that there is an increasing awareness of the role of air pollution on PV power generation, yet quantifying the correlation between anthropogenic air pollution and a reduction in solar resource using ground-based measurements seems to be lacking. This study is an attempt to provide the missing information.

II Air pollution and soiling

The work presented here only considers the reduction in solar resource due to light extinction caused by PM2.5 particles in the lower atmosphere. Air pollution causes an additional reduction in PV performance by dust accumulation on the solar panel. A discussion about the impact of different types of aerosols is given in [55], a general overview of the field is given in [56].

III Threats and Opportunities

Solar Photovoltaics are both affected by air pollution and could provide a way to reduce it. One example: approximately 2.8 billion people cook with solid fuels, rendering cooking one of the prevalent sources of fine particulate matter in India, Pakistan and Southern Africa [57]. Solar or PV powered cookers [58, 59] provide an almost pollution free alternative, and generally solar energy is thought to play an important role in improving air quality [50, 61]. However, air pollution induced scattering and extinction of light reduces the available solar resource and increase the fraction of diffuse light. These effects, if not considered correctly, can make solar powered application underperform and become unreliable [62, 63] - a threat to wide-spread adoption.

Policymakers in many regions are now acting to address air pollution issues. Dedicated Policies and the use or renewable energy sources can improve air quality, as many cities around the world have demonstrated. Notable examples include Vittoria-Gasteiz (Spain), Montréal (Canada), Lisbon (Portugal), Medellín (Colombia) and Seoul (South Korea), which managed to reduce air pollution between 28% and 63% [64]. China has drawn up an aggressive "battle plan" against smog that seems to have positive effects in cities like Beijing and Tianjin [65, 66]. In a previous action plan (2013 to 2017), a 25% reduction in PM2.5 concentration was targeted. A reduction in PM2.5 concentration is visible in **Figure 7b** for Beijing and Shanghai in the second half of 2017. Also India has formulated targets on reducing air pollution [67], though recent news do not indicate significant improvements in air quality [68].

**Summary and Conclusions:**

Urban haze is a threat with a multifaceted nature. It consists largely of anthropogenic fine particulate matter (PM2.5) generated to a large extent by incomplete combustion processes. Haze is a serious health hazard as it can penetrate deeply into the lungs and cause chronic damage to the respiratory system. Haze also affects the transition of light through the lower atmosphere, reducing the overall intensity of light reaching the ground and increasing the diffuse fraction.

*Relation between PM2.5 concentration and insolation:*

We analyze the relation between PM2.5 concentration and the loss in solar irradiance received by flat-panel silicon photovoltaic installations. Using long-term field data with high time-resolution from Delhi and Singapore, we introduce a functional relation that describes the measured data up until PM2.5 concentrations of 400 µg/m$^3$ and takes the form of a mono-exponential decay:

$$\frac{I(PM2.5)}{I_0} = exp\left(\frac{-PM2.5}{750 \pm 90}\right)$$

*Estimating insolation losses in Delhi:*

Using this function relation, we project how much insolation Delhi would have received if no air pollution was present for the period between May 2016 and November 2017. We find that within a one-year period insolation was reduced by 11.5 ± 1.5 % or 200 kWh/m$^2$ (from 1770 kWh/m$^2$ to 1570 kWh/m$^2$) by fine particulate matter.

*Projection to other cities:*

Using available air quality data and satellite measured insolation, we extended this analysis to 16 more cities in Africa (Addis Ababa, Kampala), the Americas (Bogota, Los Angeles, Mexico City), Asia (Beijing, Dhaka, Hanoi, Jakarta, Kolkata, Manama, Shanghai, Singapore, Ulan Bator) and Europe (London, Pristina). Repeating the analysis for Delhi, we found that using satellite data for a typical year (2006 – 2015) resulted in estimated losses that are consistent with those obtained from analyzing field measurements (12.2 ± 0.9%). Estimated losses range from 2.0% relative or 29 kWh/m$^2$ per year (Singapore) to 9.1% relative or 149 kWh/m$^2$ per year (Beijing). We also projected losses in PV power potential and obtained values ranging from 24 kWh/kWp (London) to 144 kWh/kWp (Ulan Bator).

*Projection to other PV technologies:*

Using spectrum measurements from Singapore, we also projected how haze affects different PV technologies. Haze affects blue light stronger than red light, resulting in PV materials with larger band gaps being more affected than silicon. Adjusting for additional current losses we predict that losses increase by 23% relative for GaAs, 33% for CdTe and 42% for perovskites (at 1.64 eV band gap) compared to silicon.

*Economic implications*

Reduced insolation results in lost revenues for system operators and investors in PV systems. Given the aggressive installation goals for many countries and communities, installations in urban areas in the order of GW can be expected. For installations of this magnitude and current electricity prices, projected annual losses in revenue for Delhi could well exceed 20 million dollars, or 12% of total revenue. Similarly, projected losses for GW installations would suffer annual losses of 16 million dollars in Kolkata and around 10 million dollars in Beijing and Shanghai. In Los Angeles with a target capacity of 1.3 GW of installed solar by 2020, projected annual economic losses are between 6 and 9 million dollars. These numbers indicate that global losses in revenue could easily amount to hundreds of millions, if not billions of dollars annually. An additional factor is that the reduced insolation and changed character of sunlight (more diffuse light), if not accounted for correctly, will affect the reliability of PV powered systems with a hard to quantify reduction in adoption.

## Acknowledgments:

The authors thank Liu Zhe for help with Chinese websites and texts. This work was financially supported by funding from Singapore's National Research Foundation through the Singapore MIT Alliance for Research and Technology's "Low energy electronic systems (SMART - LEES) IRG" and by the DOE-NSF ERF for Quantum Energy and Sustainable Solar Technologies (QESST).

**Supporting Material:**

I Typical insolation from satellite data

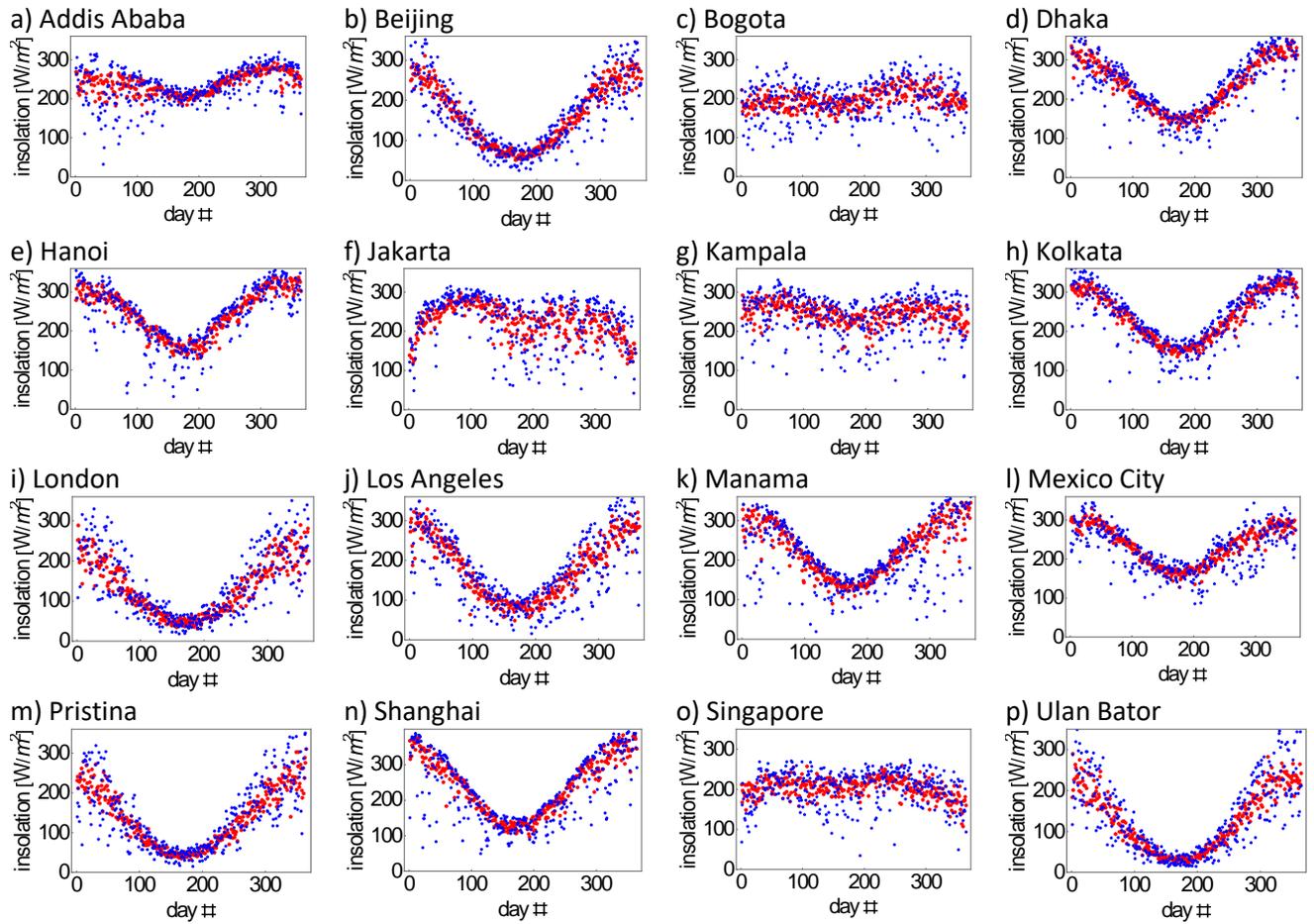

***Figure S1:*** *Median insolation based on 10 year insolation data from NASA for each day of the year and the 16 cities considered in this study (red). This data was used to estimate insolation losses due to fine particulate matter. Also shown is the insolation data for one year (2014) for comparison. Note that all given values are average insolation values for one day.*

## II Correlating PM2.5 and Insolation

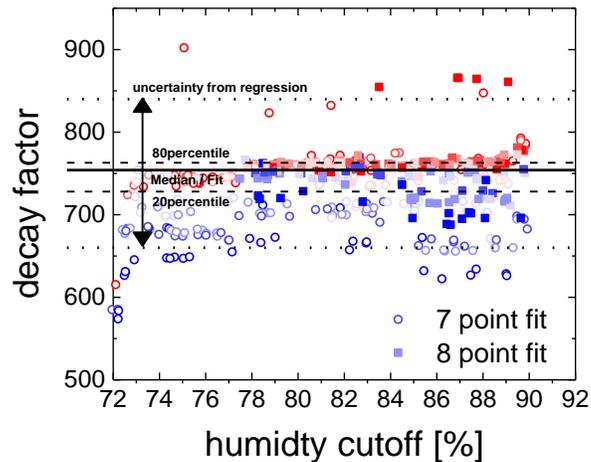

***Figure S2:*** *Monte-Carlo analysis of the exponential decay factor for varying conditions of data sorting and analysis.*

There is a choice of parameters when it comes to sorting the insolation data for different PM2.5 concentration conditions, which affects the exponential decay factor in **equation 1**. In **Figure S2** we show a Monte Carlo analysis of how the decay factor varies when several of these factors are varied randomly. The x-axis of the figure corresponds to the humidity-cutoff factor. Humidity is used as an indication for rainy days. As can be seen, no results are shown for cutoff factors beyond 90%. Including days with very high humidity results in a failure of the fitting routine. On the other hand, reducing the cutoff factor to below 78% eliminates so many data points that fitting up until a PM2.5 concentration of 400 μg/m$^3$ becomes impossible. This is indicated by the shape of the points in **Figure S2**. Rectangular shapes indicate that 8 points (i.e. concentration until 400 μg/m$^3$) were considered, whereas round shapes indicate that 7 points (until 350 μg/m$^3$) were considered. Overall, however, little dependence is found on the humidity cutoff. The value used in the analysis is 80%.

A stronger dependence can be seen on the percentile that was used to define clear-sky conditions. This filter is used to obtain a typical daily curve from the set of points at each hour and each PM2.5 concentration conditions (compare **Figure 3a**). The default value here was 0.8. Blue colors indicate a smaller value, red values a larger value. The goal here was to use a factor that was as large as possible without being influenced by outliers. As can be seen, the median value (solid line) is very close to an upper boundary beyond which only lie very few points. The default value was chose such that the result coincides with the median value. Other factors did not show a significant impact on the result.

Overall, we find that the vast majority of all points (>95%) lies within the range that the regression gave as confidence interval when fitting the linear decay for the chosen conditions. Hence, we used this confidence interval in the analysis.